\documentclass{aa}

\usepackage{graphicx}
\usepackage{txfonts}
\usepackage{color}
\newcommand{\msol}{{\rm M}_{\odot}}

\newcommand{\onvire}[1]{}
\newcommand{\eq}[1]{eq.~(\ref{#1})}
\newcommand{\beq}{\begin{equation}}
\newcommand{\eeq}{\end{equation}}

\newcommand{\remove}[1]{}

\usepackage{color}
\usepackage{epsfig}
\usepackage{amssymb}
\usepackage{graphicx,natbib}
\begin{document}
  \title{Dead zones in protostellar discs: the case of Jet Emitting Discs} 

  \author{C. Combet
         \inst{1}
         \and
         J. Ferreira\inst{2}
         \and F. Casse\inst{3}
         }

  \offprints{celine.combet@astro.le.ac.uk}

  \institute{Dept. of Physics and Astronomy, University of Leicester,
    Leicester, LE1 7RH, United Kingdom \\ \and Laboratoire d'Astrophysique de Grenoble, UJF/CNRS,
             BP 53, 38041 Grenoble Cedex 9, France \\ \and Laboratoire APC, Universit\'e Paris Diderot, 10, rue A. Domon et L. Duquet 75205, Paris Cedex 13, France}

  \date{Received \today; accepted }


 \abstract
  {Planet formation and migration in accretion discs is a very active
    topic. Among the many aspects related to that question, dead zones are of
  particular importance as they can influence both the formation and the
  migration of planetary embryos. The ionisation level in the disc is the key 
  element in determining the existence and the location of the dead zone. This has been studied either within the Standard Accretion Disc (SAD) framework or using parametric discs.}
  {In this paper, we extend this study to the case of Jet Emitting
    Discs (JED), the structure of which strongly differ from SADs because of
    the new energy balance and angular momentum extraction imposed by the jets.  }
  {We make use of the ($r,z$) density distributions provided by self-similar accretion-ejection models,
    along with the JED thermal structure derived in a previous paper, to create maps of
    the ionisation structure of JEDs. We compare the ionisation rates we obtain to the critical value required to trigger the magneto-rotational instability.}
  {It is found that JEDs have a much higher ionisation degree than SADs which 
renders very unlikely the presence of a dead zone in these discs. }
  {As JEDs are believed to occupy the inner regions of accretion discs, the extension of the
dead zones published in the literature should be re-considered for systems in
which a jet is present. Moreover, since JEDs require large scale magnetic fields close to equipartition, our findings raise again the question of magnetic field advection in circumstellar accretion discs.}

\authorrunning{Combet et al.}
\titlerunning{Dead zones and Jet Emitting Discs}

  \maketitle
%

\section{Introduction}

A detailed knowledge of the dynamics and chemistry of protostellar accretion discs
is unavoidable if one is to understand the initial conditions of planet 
formation and the environment in which their migration proceeds. 
The theory of protostellar accretion discs has been widely developed since the seminal work of
\citet{1973A&A....24..337S} and is commonly set in the framework of the standard accretion 
disc (SAD) in which viscosity is the sole agent to remove angular momentum and allow 
accretion (e.g. \citealp{1981ARA&A..19..137P, 1998ApJ...500..411D,2005A&A...442..703H}). 
The current paradigm is that this is done through MHD turbulence, triggered and 
sustained in the disc by the magneto-rotational instability 
(hereafter MRI, \citealt{1991ApJ...376..214B}). 

This instability requires a good coupling between 
the disc material (mostly neutrals) and the magnetic field, therefore requires a
minimum degree of ionisation to exist. If the ionisation level is not high enough 
for the MRI to exist, the viscosity is very low: this results in a very low level of accretion 
in the corresponding region of the disc, then termed ``dead zone'' \citep{1996ApJ...457..355G}.
Dead zones may play an  important role in the field of planet formation and migration.
Rossby vortices can be triggered at the interface between an active and a dead zone, trapping
solids within them and leading to a burst of planet formation 
\citep{2006A&A...446L..13V,2009A&A...497..869L}.
They also have been put forward as an agent able to halt planetary migration, a long-standing
issue in explaining planetary systems such as our own \citep{2006ApJ...642..478M, 2009ApJ...691.1764M}.
Locating dead zones in an accretion disc is therefore an important task.
Several authors have studied this question, generally using either i) a passive disc 
(\citet{1997ApJ...480..344G} uses a power-law disc, \citet{2003ApJ...598..645M} 
the passive disc from \citet{1997ApJ...490..368C}), or ii) a standard accretion disc 
\citep{2002MNRAS.329...18F}. These studies show that a dead zone can generally exist
in the inner parts of accretion discs, typically extending from 0.1--1~AU up to a few AU. 

Another way to indirectly probe the physics of the inner disc regions is to look at the jets from these sources. Bipolar jets are fundamental to star formation and are observed from the earliest (Class 0) to the latest stages (classical TTauri stars, Class
2) of the process. They are believed to originate from the inner disc (up to a
few AU), precisely where the aforementioned works predict the existence of a
dead zone. Extended MHD disc winds are, to date, the best candidates to
explain jet kinematics \citep{2006A&A...453..785F}. Because they change the energy 
balance of the disc, the jet emitting disc (JED) structure differs from that of the SAD one.
This was shown in \citet{2008A&A...479..481C} (paper~I hereafter) and questions the existence dead zones in the inner parts of jet launching discs. This is the issue we address in the present paper which is outlined as follows.
We start by a quick recap of disc wind solutions  and JED structures in section~\ref{sec:jed_recap}.
Section~\ref{sec:ion_calcul} presents the procedure we follow to calculate the ionisation 
level of a JED while the results are shown in section~\ref{sec:results}. Finally, 
discussion and conclusions are drawn in section~\ref{sec:discussion}.

\section{Jet Emitting Discs: a quick reminder\label{sec:jed_recap}}

It is widely accepted that protostellar jets observed in young stellar objects (YSOs)
are launched from the surface of accretion discs \emph{via} the 
magneto-centrifugal launching mechanism, originally developed by
\citet{1982MNRAS.199..883B}. Solutions to the problem have generally been
found by treating the disc as a boundary condition \citep{1982MNRAS.199..883B,2000MNRAS.318..417V},
 hence preventing any quantification of the interplay between the disc and the jet. 
This difficulty has been lifted in the self-similar framework of the Magnetised
Accretion-Ejection Structures (MAES) model, where the set of steady-state MHD equations are
consistently solved, going from the resistive regime within the disc to the ideal
MHD wind/jet \citep{1995A&A...295..807F,1997A&A...319..340F,2000A&A...353.1115C,2004ApJ...601L.139F}.
MAES rely on a self-sustained turbulence to provide 
the required anomalous transport of magnetic field (through an effective magnetic diffusivity),
along with magnetic fields smaller than, but close to, equipartition (i.e., the plasma $\beta=2\mu_0 P/B^2 \gtrsim 1$). 
This may appear problematic as the magneto-rotational instability, invoked as the source of turbulence, 
is known to be quenched whenever the field reaches equipartition. This difficulty is lifted when remembering that 
(1) MAES are close to but always below the marginal stability limit for the MRI \citep{1991ApJ...376..214B, 1994MNRAS.270..138G}
(2) the MRI could still persist at equipartition \citep{2005A&A...444..337B}.
Note also that other sources of MHD turbulence could be at work around equipartition in a medium as complex as a stratified accretion disc 
\citep{2002ApJ...569L.121K}.


\subsection{Self-similar MAES solutions}

In the present study we use a disc wind solution undergoing a surface heating described in \citet{2000A&A...361.1178C}. 
Due to this heating, the vertical plasma pressure is increased which enhances substantially the jet mass load. 
This additional heating was assumed to be powered by dissipation of MHD disc turbulence and
parametrised as a tiny fraction of the disc accretion power ($f = 8 \times 
10^{-4}$ in the solution chosen here). This new class of solutions,
referred to as ``warm'' in \citet{2004A&A...416L...9P}, is still ``cold'' in a
dynamical sense as the initial thermal energy remains negligible
compared to gravity, and wind acceleration is still mostly
magnetic. The disc ejection efficiency $\xi$ is defined by $\dot M_{\rm acc} \propto r^\xi$. While isothermal disc wind models 
provide $\xi$ around 0.01, 
these warm models can achieve larger values. For the solution used here, we have $\xi= 0.04$ from a disc of aspect ratio $\epsilon=h/r= 0.03$, 
where $h(r)$ is the disc half thickness. If the JED is established between say, $r_{\rm in}= 0.04$ AU and 
$r_{\rm J}= 0.3 -1$~AU\footnote{The question of the radial extension of the launching region of the disc 
has been debated at length. If X-wind configurations, where the jet comes from
the innermost region of the disc \citep{2000prpl.conf..789S}, could surely
contribute to the mass loss, an extended disc-wind component appears indispensable, especially in the younger Class~I 
objects \citep{2006A&A...453..785F}.}, this translates into an ejection to accretion mass rates ratio 
$2 \dot M_{\rm jet}/\dot M_{\rm acc}\simeq \xi \ln \frac{r_{\rm J}}{r_{\rm in}}= 0.08$ to $0.13$,
in agreement with observations. The asymptotic poloidal jet velocity is a factor $\sqrt{2\lambda-3}$ 
larger than the keplerian velocity at the field line footpoint, 
where $\lambda \simeq 1 + 1/2\xi$ is a measure of the magnetic lever arm  \citep{1997A&A...319..340F}. 
With $\lambda=13.15$, this model is found to reproduce 
well several observed properties of atomic T Tauri jets such as the rotation signatures, the ejection 
to accretion ratio, and the centroid velocities in 
atomic spectra along the jet axis (e.g. \citealt{2004A&A...416L...9P,2007IAUS..243..203C}).

\begin{figure}
\begin{center}
\includegraphics[clip=,width=\columnwidth]{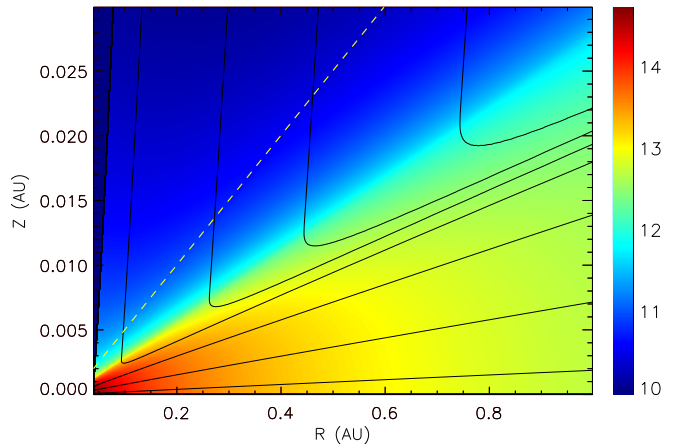}
\caption{Log-density (cm$^{-3}$, colour contours) and streamlines (solid black
 lines) of a JED powering jets that seem consistent with observations (see text for more details). The dashed yellow line shows the position of 
the slow magnetosonic (SM) point (located at $z_{SM}=1.67 h$). The SM point occurs in the ideal MHD regime above the disc, 
whereas the flow is deflected vertically (ie $u_r=0$) below, in the underlying resistive layers. 
Note that in order to resolve the disc internal structure, the vertical scale has been expanded.
\label{fig:maes}
}
\end{center}
\end{figure}

The resulting ($r,z$) distribution of density is shown in fig.~\ref{fig:maes} along with some streamlines 
(black solid lines). 
Note that the range of launch radii produces an ``onion-like'' structure with a fast, dense axial beam 
surrounded by progressively slower and wider streamlines that recollimate on larger scale, producing a high apparent jet collimation. 
This is however not shown here as this work focuses on the properties of the Jet Emitting Disc only. As can be seen on the figure, 
only the upper layers of the disc are deviated into the outflow, while most of the material ends up at the inner boundary of the disc, 
to be accreted by the protostar.

The global picture of a protostellar accretion disc would then be the following
(see fig.1 of paper~I):
i) in the inner parts from ($r_{\rm in}$ to $r_{\rm J}$), a jet emitting disc (JED) 
having MAES properties and ii) a standard
accretion disc (SAD) from $r_{\rm J}$ to $r_{\rm  out}$. As mentioned above, SADs
rely solely on the radial viscous transport of angular momentum for accretion
to proceed. Conversely, MAES solutions show that as soon as a powerful jet is present, most of the
angular momentum and energy within the disc are evacuated vertically in
jets. This suggests that JEDs should have a very different structure than SADs
and is developed hereafter.

\subsection{The jet emitting disc's structure}

In paper~I, we analytically derived the radial properties
of JEDs for a given protostellar mass $M$ and mass accretion rate in the disc
$\dot M_{\rm acc}$. This was done in a 
formalism analogous to that of SADs, but using the energetic balance suggested
by the MAES solutions. It was found that for a given $M-\dot M_{\rm acc}$ combination, a JED
is cooler, lighter and thinner than its equivalent SAD (we refer the reader to
that paper for more details).

\begin{figure}
\begin{center}
\includegraphics[clip=,width=\columnwidth]{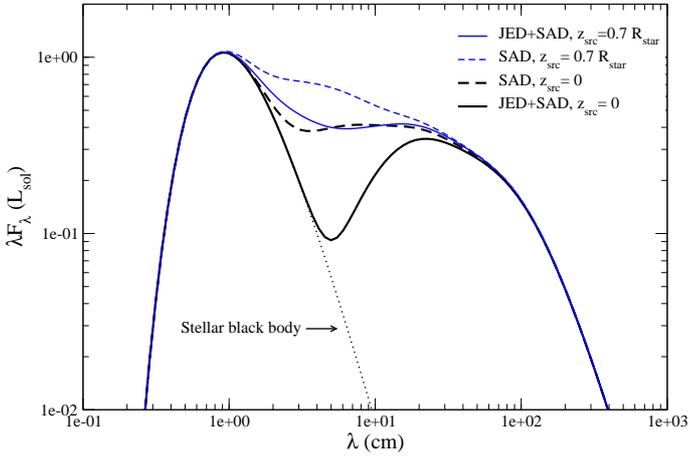}
\caption{SED for different disc configurations with $\dot M_{\rm acc}=10^{-7}\;\msol$~yr$^{-1}$. Solid lines correspond to a
 disc composed of a JED up to $r_J=0.3$~AU and a SAD in the outer regions.
Dashed line represent the case where the SAD fills all the disc, up to the
innermost radius $r_{\rm in}=0.04$~AU. For each case, the source of
irradiation (UV from accretion column hot spot) had been located in $z=0$
(thick lines) and $z=0.7R_{\rm star}$ (thin lines).\label{fig:compare_sed}}
\end{center}

\end{figure}

Its lower temperature should allow for the presence of a JED to
be tested by the spectral energy distribution (SED) of the
disc. This was illustrated using very crude assumptions\footnote{In fig.~4 of paper~I 
we calculated the SED assuming a given incident angle for the radiation. This
lead to an overestimation of the irradiation flux in the inner parts of the
disc making it too bright at small wavelength. This is corrected here by
using the calculated JED geometry consistently to estimate the incident angle
at each radius. The qualitative conclusions of paper~I are
unchanged, simply the actual values of the flux.} for the disc geometry in
paper~I. Therefore, we present an updated
version of the SED in
fig.~\ref{fig:compare_sed} for illustrative purposes. Both viscous heating and illumination from the
central object are included in the SED calculation. Assuming most of the
irradiation comes from the hot spot of the accretion column onto the protostar,
we calculate the SED for two altitudes of the hot spot, $z_{\rm src}=0$ and $z_{\rm src}=0.7R_\star$.
We consider both the cases where the SAD extends down to $r_{\rm in}$ and
where it is supplemented by a JED in the inner parts, with $r_J=0.3$~AU. As can
be seen in fig.~\ref{fig:compare_sed}, for a given spot altitude, there
is a clear signature of the presence of the JED. However, it appears very
hard to distinguish between a JED+SAD configuration with $z_{\rm src}=0.7R_\star$ and
a single SAD with $z_{\rm src}=0$. This simple example illustrates that one
cannot solely rely on the SED to test for the presence of a JED in
a given system. But this indirect method should nevertheless prove useful when used
in combination with other diagnostics (e.g. jet kinematics,  \citealp{2006A&A...453..785F}).

\section{Calculating the disc ionisation\label{sec:ion_calcul}}

Several agents can contribute to the ionisation of an accretion
disc. In his seminal work on dead zones, \citet{1996ApJ...457..355G} considered ionisation 
by the cosmic ray background only, of a power-law disc. Using a passive disc,
\citet{2003ApJ...598..645M} included in their calculations ionisation from cosmic rays, X-rays and radioactive decays (that remains marginal) in the disc. They concluded that cosmic ray (CRs) ionisation was the leading process while the X-ray contribution could become dominant if the X-ray photon energy
was sufficient ($5-10$ keV). This result certainly holds if CRs can reach the
disc. However it has been argued that they may not be able to do so because of
the winds associated to star formation, in which case, X-ray ionisation
becomes fundamental again
\citep{1997ApJ...480..344G,2002MNRAS.329...18F}. This is particularly true
for JEDs that are, by definition, the source of a strong wind that eventually
recollimates into a jet. Therefore, we only focus in this paper on the X-ray ionisation of
the JEDs. 

\subsection{Ionisation rate from X-rays}

A well known result from X-ray ionisation studies is that the ionisation rate
is dominated by secondary ionisation, i.e. the ion pairs collisionnaly  produced by a fast
primary photo-electron \citep{1983ApJ...267..610K}.
The X-ray ionisation rate is written as (\citealp{2003ApJ...598..645M} and
references therein)

\beq
\zeta_X= \frac{L_X}{4\pi d^2 k T_X}\; \sigma(k T_X)\; \left(\frac
    {kT_X}{\Delta \epsilon}\right)\; J(x_0, \tau)
\label{eq:zeta_x}
\eeq
with
\beq
J(x_0, \tau)=\int_{x_0}^{\infty} x^{-n} \exp\;\left[-x-\tau(kT_x)x^{-n}\right]
dx\;,
\label{eq:attenuation}
\eeq
where $L_X $ is the source X-ray luminosity, $d$ is the distance from the
source, $kT_X$ the energy of the photons, $\sigma(k T_X)$ is the absorption
cross section, $\Delta \epsilon \approx 37$~eV is the average energy required for
ion pair production\footnote{The parenthesis should read
 $(kT_X-IP)/\Delta\epsilon$, as the energy of the primary photo-electron
 depends on the ionisation potential ($IP$) of the medium. However, $IP$ is
 $\sim$ a few eV whereas $kT_x\gtrsim 1$~keV.} and $J(x_0, \tau)$ is the attenuation factor. The latter
is calculated using $x=E/kT_X$ and is function of the optical depth,
$\tau(k T_X)= N_H \sigma (kT_X)$ with $N_H$ the column density along the line
of sight, and of the cut-off energy $x_0=E_0/kT_X$. The absorption cross
section is energy dependent and expressed as
\[
\sigma (k T_X)= \sigma_0 \left(\frac{kT_X}{1 {\rm keV}}\right)^{-n}
\]
with $\sigma_0=8.5\times 10^{-23}$~cm$^2$ and $n=2.81$
\citep{1997ApJ...480..344G,1999ApJ...518..848I, 2002MNRAS.329...18F}.

\subsection{Recombination processes \label{subsec:dust}}
Electrons in an accretion disk may recombinate \emph{via} several channels, by interacting 
with both the gas and dust grain phases. Dissociative recombinations in the gas phase 
have been extensively studied both theoretically and experimentally, giving a relatively well-defined 
picture of recombination with molecular or metallic ions (for a review, see \citealp{2006PhR...430..277F}). 

The picture becomes however far more complicated when gas-grain chemistry is considered. Several authors 
have investigated the importance that recombinations of electrons on the surface of dust grains 
may play on the ionisation level of accretion discs 
\citep{2000ApJ...543..486S,2004A&A...417...93S,2006A&A...445..205I, 2007Ap&SS.311...35W,2008MNRAS.388.1223S}.
It is generally found that including grain chemistry decreases the ionisation level of the disc (with respect to the
results of gas-phase calculations only), provided that 
the grains are small enough ($\lesssim 1\;\mu{\rm m}$) and well-mixed with the gas, on the
whole disc thickness \citep{2000ApJ...543..486S,2006A&A...445..205I}. These conditions may not always be fulfilled 
as theoretical considerations and SEDs of TTauri accretion disks suggest that grains 
grow and settle onto the disc midplane during the protostellar evolution \citep{2004A&A...421.1075D,2006ApJ...638..314D}.
Note also that in most of the models in \citet{2006A&A...445..205I}, the gas-only and gas-grain chemistry 
converge in the inner disc ($\lesssim 0.6$~AU), where dust grains may not survived. 
Finally, the detailed numerical studies investigating the effect of dust grains on the MRI 
\citep{2008MNRAS.388.1223S,2010ApJ...708..188T}, generally do so at distances of a few AU that are not relevant to the JED.

Given the complexity and uncertainty surrounding the grain chemistry and distribution in the discs, 
we restrict ourselves to recombination processes in the gas-phase for the remainder of this paper.
This choice is motivated by the fact that JED are relevant only to the innermost regions of the discs,
where dust grains may not survive and where it is difficult to extrapolate the results of 
existing studies.

\subsection{Ionisation fraction}

The ionisation fraction $x_e=n_e/n_H$ is generically obtained from the solution of the evolution equation of the electronic density $n_e$. Considering the gaseous phase only, this evolution is described by system of coupled equations 
(e.g. \citealp{2002MNRAS.329...18F})
\begin{eqnarray}
\label{eq:evolution_ne}
\frac{dn_e}{dt} & = & \zeta_X n_H - \beta n_en_{{\rm m}^+}-\beta_r n_e n_{{\rm M}^+}\\\nonumber
\frac{dn_{{\rm m}^+}}{dt} & = & \zeta_X n_H - \beta n_en_{{\rm m}^+}-\beta_t
n_{M}n_{{\rm m}^+}\;,
\end{eqnarray}
where $n_M$ and $n_{{\rm M}^+}$ are the density of metals and metal ions
respectively, $n_{{\rm m}^+}$ is the density of molecular ions, the ionisation
rate $\zeta_X$ is given by \eq{eq:zeta_x}, $\beta$ is the dissociative
recombination rate, $\beta_r$ is the radiative recombination rate and
$\beta_t$ is the rate of charge transfer between molecular ions and metal
atoms.
\remove{Assuming there are no metals ($n_{M}=n_{{\rm M}^+}=0,\; n_e=n_{{\rm m}^+}$), these equations reduce to
\beq
\frac{dn_e}{dt}  =  \zeta_X n_H - \beta n_e^2.
\eeq
}

Steady-state, that is assumed in a majority of the studies published in the
subject, is valid if the ionisation and recombination timescales are shorter than the accretion
timescale in the disc. This is true for SADs as they are characterised by a slow
inward radial velocity, hence long accretion timescales. Also assuming no
metals  ($n_{M}=n_{{\rm M}^+}=0,\; n_e=n_{{\rm m}^+}$),
\eq{eq:evolution_ne} leads to
\beq
x_e=\sqrt{\frac{\zeta_X}{\beta n_H}}\;.
\label{eq:xe}
\eeq
A priori, more caution is necessary in the case of JEDs where the
accretion proceeds much faster, namely at sonic velocities. For a JED, the accretion timescale
is therefore given by 
\beq
t_{\rm acc}\sim \frac{r}{c_s}\;\;(c_s\;\;\textrm{is the isothermal sound
 speed})
\label{eq:tacc}
\eeq
while the ionisation and recombination timescales obtained
from \eq{eq:evolution_ne} read
\beq
t_{\rm ion}=\frac{x_e}{\zeta_X}
\label{eq:tion}
\eeq
and
\beq
t_{\rm rec}=\frac{1}{\beta x_e n_H}\;.
\label{eq:trec}
\eeq
In steady-state, these two latter timescales are equal, $t_{\rm ion}=t_{\rm rec}$.
For our calculations, we first assume that steady-state holds and 
we check \emph{a posteriori} (in \S\ref{subsec:timescales}) 
the extent to which this hypothesis is valid.

\subsection{Dead zones and critical ionisation rate \label{subsec:critical_ioni_rate}}

The MRI, that arguably provides the disc's anomalous viscosity and resistivity \citep{2009A&A...504..309L, 2009ApJ...697.1901G}, only sets in if the
Alfv\'enic time scale $\tau_A = h/V_A$ is shorter than the diffusion time scale $\tau_d = h^2/\eta$ due to electron-ions collisions.
Given the Alfv\'en velocity of the medium $V_A$ and the disc scale height $h$,
this criterion translates into a magnetic Reynolds number ${\cal R}_m = h
V_A/\eta$ that must be greater than a critical value ${\cal R}_m^{\rm crit}$. The
value of the critical Reynolds number is known to depend on various factors, such as
the presence of a net magnetic flux through the disc, but is 
not firmly established yet. Therefore, we adopt the approach widely found in the 
literature and consider both ${\cal R}_m^{\rm crit}=1$ and  ${\cal R}_m^{\rm crit}=100$. 
Using the Ohmic resistivity
\citep{1994ApJ...421..163B}
\[
\eta=234\;\frac{T^{1/2}}{x_e}\; \textrm{cm}^2\textrm{s}^{-1}\;,
\]
and the fact that a JED requires a magnetic field close to 
equipartition \citep{1995A&A...295..807F}, this condition becomes\footnote{\citet{1996ApJ...457..355G} gives an equivalent expression of
 ${\cal R}_m$   for a SAD using \citet{1981PThPS..70...35H}'s
resistivity and the central disc temperature $T_o$ instead of the disc aspect ratio $\varepsilon$.}
\beq
{\cal R}_m \sim 10^{13} x_e
\left(\frac{\varepsilon}{0.01}\right)\left(\frac{r}{1 {\rm AU}}\right)\; \gtrsim {\cal R}_m^{\rm crit}\;,
\label{eq:rm}
\eeq
where $\varepsilon=h/r$ is the disc aspect ratio. From \eq{eq:xe} and
using the disc vertical hydrostatic equilibrium, the definition of the
accretion rate and the collisional recombination
rate $\beta= 3\times 10^{-6}T^{-1/2}\;{\rm cm}^{3}\;{\rm s}^{-1}$ (e.g. \citealp{2002MNRAS.329...18F}) one gets the ionisation fraction to be
\begin{eqnarray}
\nonumber
x_e =&1.5\times 10^{-12}&\left(\frac{\zeta_X}{10^{-17}{\rm
   s}}\right)^{1/2}\left(\frac{\varepsilon}{0.01}\right)^{3/2}\\
&\times&\left(\frac{M}{\msol}\right)^{1/2}\left(\frac{\dot M}{10^{-7}\msol{\rm
   yr}^{-1}}\right)^{-1/2}\left(\frac{r}{1 {\rm AU}}\right)^{1/2}\;.
\label{eq:xe_crit}
\end{eqnarray}
Inverting \eq{eq:rm}, this leads to a critical ionisation rate

\begin{eqnarray}
\nonumber
\zeta_X^{\rm crit}= &4.3\times 10^{-20}\; \left({\cal R}_m^{\rm crit}\right)^2&\left(\frac{\varepsilon}{0.01}\right)^{-5}
\left(\frac{M}{\msol}\right)^{-1}\left(\frac{\dot M}{10^{-7}\msol{\rm
   yr}^{-1}}\right)\\
&\times &\left(\frac{r}{1 {\rm AU}}\right)^{-3}\;{\rm s}^{-1}\;.
\label{eq:zeta_crit}
\end{eqnarray}
Note that this result is obtained for the typical recombination rate
 $\beta$ mentioned above. 

\subsection{X-ray properties of young stellar objects}

Young stellar objects of all stages (Class 0 to III) are known to be X-ray
emitters. In the 1980's, the first discoveries were made by the 
\emph{Einstein} X-ray observatory in the less embedded objects 
(Class II and III, e.g. \citealp{1981ApJ...243L..89F}). 
The ASCA and ROSAT satellites allowed the detection of X-rays from Class~I protostars during the
following decade (e.g., \citealp{1996PASJ...48L..87K,
 1997Natur.387...56G}). More recently, the \emph{Chandra} Orion ultra-deep
project (COUP) survey, revealed X-ray detection in the youngest Class~0
objects (e.g. \citealp{2008ApJ...677..401P}).

YSOs typical X-ray luminosities range between $10^{28}-10^{30}$~erg~s$^{-1}$,
with flares reaching $10^{32}$~erg~s$^{-1}$ and with typical photon
temperatures of $kT_X=1-5$~keV. 
Quiescence properties may vary slightly from one class of object to another but 
typical values are $L_X^{\rm quiesc}=10^{29}$~erg~s$^{-1}$ and $kT_X^{\rm quiesc}=1-3$~keV
\citep{2003PASJ...55..653I,2005ApJS..160..469F,2005ApJS..160..423W}.

Using the statistics of the COUP survey, \citet{2005ApJS..160..423W} determined a rate 
of one flare per week per star, with flare durations ranging from a few hours up to 3 days.
The flare luminosities, spectral
properties and variability  distributions appear fairly constant
through all the stages of protostellar evolution, suggesting single underlying mechanism
\citep{2008ApJ...677..401P}. Fiducial flare characteristics are 
$L_X^{\rm flare}=10^{30-31}$~erg~s$^{-1}$ and a spectrum harder than in quiescence, 
$kT_X^{\rm flare}\gtrsim 5$~keV.  

One might expect these flares to be due to reconnection events in magnetic loops 
connecting the protostar to its circumstellar disc. However,  \citet{2008ApJ...688..437G} found
no differences between accreting and disc-free systems (i.e. Class~II and Class~III). Thus, these reconnection events 
must be related to magnetic loops in the stellar magnetosphere alone.
The size of these loops, hence the location of the X-ray emitter, can be 
estimated from the flare properties but remain ``flare-model'' dependent. From 
the COUP survey, \citet{2008ApJ...688..418G} obtains half-lengths of
$L\sim10^{11}-10^{12}$~cm. 

From the above, we see that the X-ray source alone introduces three parameters in the problem:
i) its luminosity $L_X$, ii) the photon temperature $kT_X$ and iii) its
location/altitude $z_{\rm src}$, linked to the magnetic tube length. 
These three parameters will be systematically 
varied within their (reasonable) respective ranges.

\subsection{Choosing the appropriate self-similar solution}

Using \eq{eq:zeta_x}, we wish to calculate the ionisation rate of a JED for a
given set of $M-\dot M_{\rm acc}$. To that end, the appropriate MAES solution
needs to be selected. On the one hand, each $M-\dot M_{\rm
 acc}$ gives rise to a unique JED aspect ratio profile $\epsilon(r)$, as shown in
paper~I (see also fig.~\ref{fig:epsilon_classes}). Such a profile is in particular strongly  dependent on the dominant opacity regime, which varies along the radius. On the other hand, a self-similar MAES solution
is characterised by a specific value of $\epsilon$. Therefore, for a given system
($M-\dot M_{\rm acc}$), we have to select a self-similar solution whose $\epsilon$
best matches that of the JED radial structure\footnote{There naturally is  a degeneracy 
in $M-\dot M_{\rm acc}$, several combinations of which giving rise to approximately
the same values of $\epsilon$. However, this is not a point we wish to explore
in the present paper.}. 

Our system's canonical physical parameters are inspired from the Class I/II
object DG~Tau and read $M=0.5\;\msol$ and $\dot M_{\rm
 acc}=10^{-6}\;\msol$~yr$^{-1}$. For these values the JED radial profile calculation gives an average value over the JED extension of  $\epsilon \sim 0.03$ (Paper I). As described in Paper~I, this calculation is
valid in the optically thick regime which extends, for these particular
parameters, up to $r \sim 1.8$~AU.
%
%
The MAES solution shown in fig.~\ref{fig:maes} is characterised by
$\epsilon=0.03$ and is the reference solution that will be considered in the
following section. We also arbitrarily choose a radial extension of $r_J=1$~AU only, well 
inside the validity region of the JED calculation.

\section{Results \label{sec:results}}

Before exploring the question of the existence or absence of a dead zone in
JEDs (\S\ref{subsec:jed_ioni}), we first present the validity range of the
steady-state hypothesis we make.

\subsection{Timescales and
 steady-state \label{subsec:timescales} \label{subsec:ss}}

To infer the ionisation fraction from the ionisation rate using \eq{eq:xe}, steady-state
is assumed. Once the ionisation rate and fraction are known, we can verify
\emph{a posteriori} whether or not this hypothesis holds ($t_{\rm
 ion} < t_{\rm acc}$), using \eq{eq:tacc} and \eq{eq:tion}. Although this check
can only be done once the calculations in \S\ref{subsec:jed_ioni} are
performed, it seems more relevant to present these considerations beforehand.

As evoked by fig.~\ref{fig:maes}, the problem
should be treated in a full 2D configuration. However, we present several
results at the disc midplane. Indeed, it is the hardest part of the
disc to ionise because of the amount of material the X-rays have to
cross, and is therefore the most constraining location to look at.

\begin{figure}
\includegraphics[clip=,width=\columnwidth]{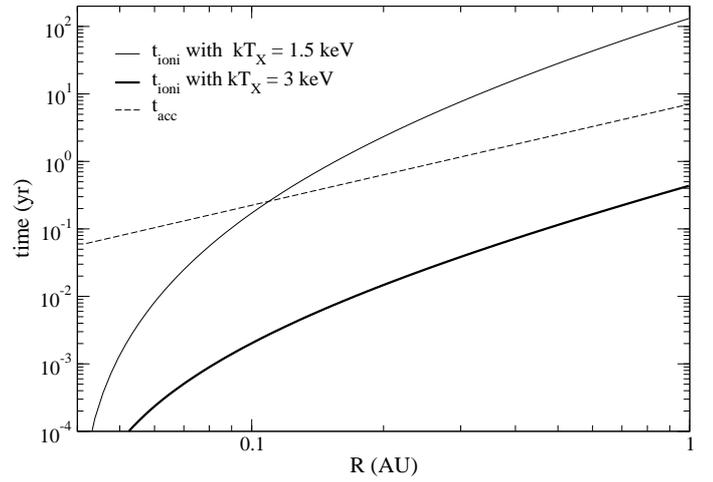}
\caption{Accretion (dashed) and ionisation (solid) timescales
 on the disc midplane as function of distance from the central object, with
 the X-ray source ($L_X=10^{29}$~erg~s$^{-1}$) located on the midplane ($z_{\rm src}=0$). Two
 electron temperatures have been considered. Calculations done for our reference MAES solution.
 \label{fig:timescales}}
\end{figure}

The radial evolution of the accretion and midplane ionisation timescales are plotted
in fig.~\ref{fig:timescales}. When the system's physical properties are fixed ($M$, $\dot M$, $\epsilon$),
the accretion time (dashed line) simply scales as $t_{\rm acc} \propto r^{3/2}$. The
solid lines represent the evolution of the ionisation timescale, for a source
located in $z_{\rm src}=0$ with $L_X=10^{29}$~erg~s$^{-1}$. For a 3~keV photon
(thick solid line), the ionisation time is always shorter than the dynamical one which validates
the use of the steady-state for this configuration. This is not the
case for the 1.5~keV photon (thin solid line) where the ionisation time is
longer than the accretion time when $r\gtrsim 0.1$~AU. This is just an example
illustrating a situation where the full time-dependent treatment should be
used.

\begin{figure}
\includegraphics[clip=,width=\columnwidth]{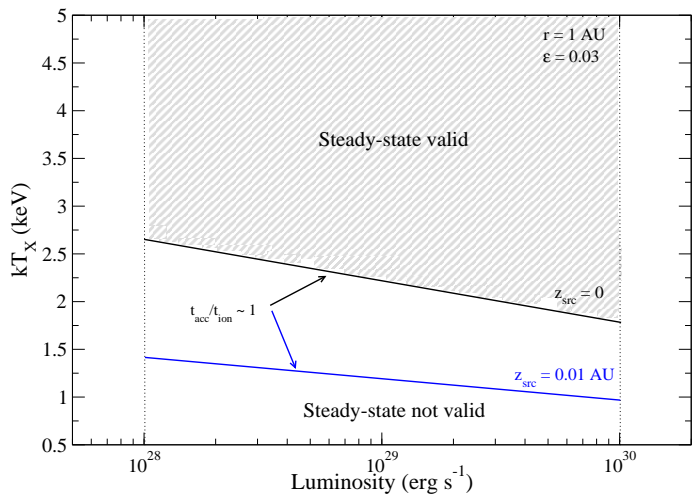}
\caption{Black and blue solid lines represent $t_{\rm acc}=t_{\rm ion}$ in the luminosity-temperature plane for an X-ray source located at $z_{\rm src}=0$ and $z_{\rm
   src}=3\;R_\odot$ respectively. Steady-state is valid only if $t_{\rm acc} > t_{\rm ion}$, namely above each line. The shaded area provides the most conservative estimate. Calculations are done with our reference MAES solution.    
\label{fig:ss_valid}} 
\end{figure}

Figure~\ref{fig:ss_valid} summarises the validity regime of the steady-state
hypothesis at  $r=1$~AU (again, more constraining than any smaller radii). In the luminosity-photon energy plane, the solid black and
blue lines locate where $t_{\rm acc}=t_{\rm ion}$ for $z_{\rm src}=0$ and
$z_{\rm src}=3 R_\odot$ respectively. Above the line, steady-state can be assumed
whereas it fails below it. The frontier between the two regimes naturally
changes with both luminosity and photon energy: a high luminosity compensate
for a small photon temperature and \emph{vice versa}. The source
altitude is also a critical parameter: most of the parameter space is covered by the steady-state
hypothesis for $z_{\rm src}=3 R_\odot$ (blue line), whereas it is not the case
for $z_{\rm src}=0$ (black line). For the latter, a time dependent approach should be used
for the lowest photon temperatures ($kT_X\lesssim 2$~keV).

The validity of the steady-state hypothesis depends on the X-ray source parameters (luminosity, energy,
location). Overall, this approach holds on most of the parameter space.
In the following we restrict ourselves to steady-state calculations, keeping
in mind they break down for unfavourable combinations of the parameters.

\subsection{Ionisation of JEDs\label{subsec:jed_ioni}}

\subsubsection{Ionisation rates}

The ionisation rate of the MAES solution presented in fig.~\ref{fig:maes} is calculated
using \eq{eq:zeta_x}. A $500 \times 500$ MAES box ensures the convergence
of the column density numerical integration necessary for the optical depth calculation.
The attenuation factor $J(x_0, \tau)$ given by \eq{eq:attenuation} is computed
from $x_0=1 {\rm keV} / kT_X$ to $x_{\rm max}=100$ \citep{2003ApJ...598..645M}.

\begin{figure*}
\begin{center}
\includegraphics[clip=,width=\columnwidth]{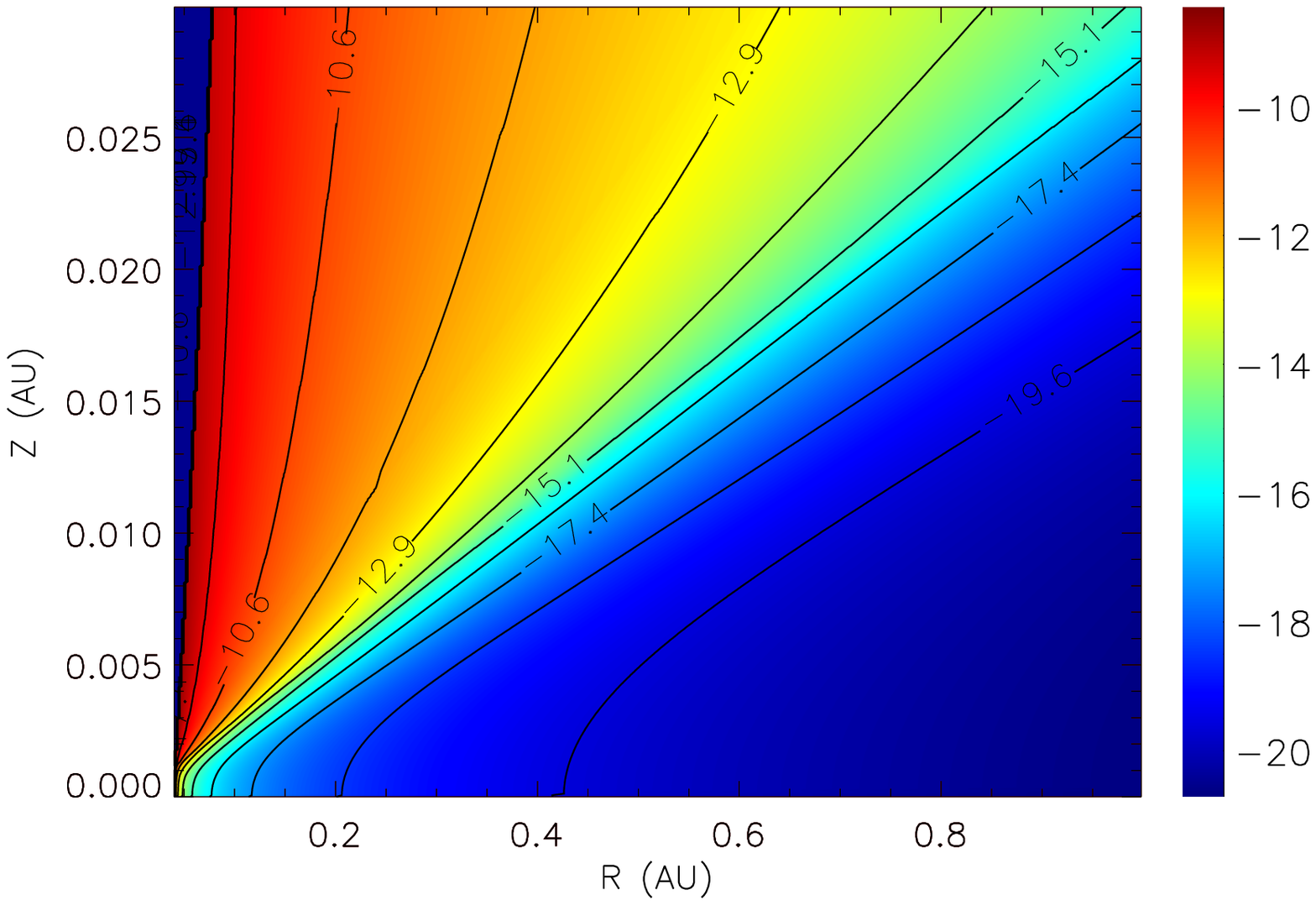}
\includegraphics[clip=,width=\columnwidth]{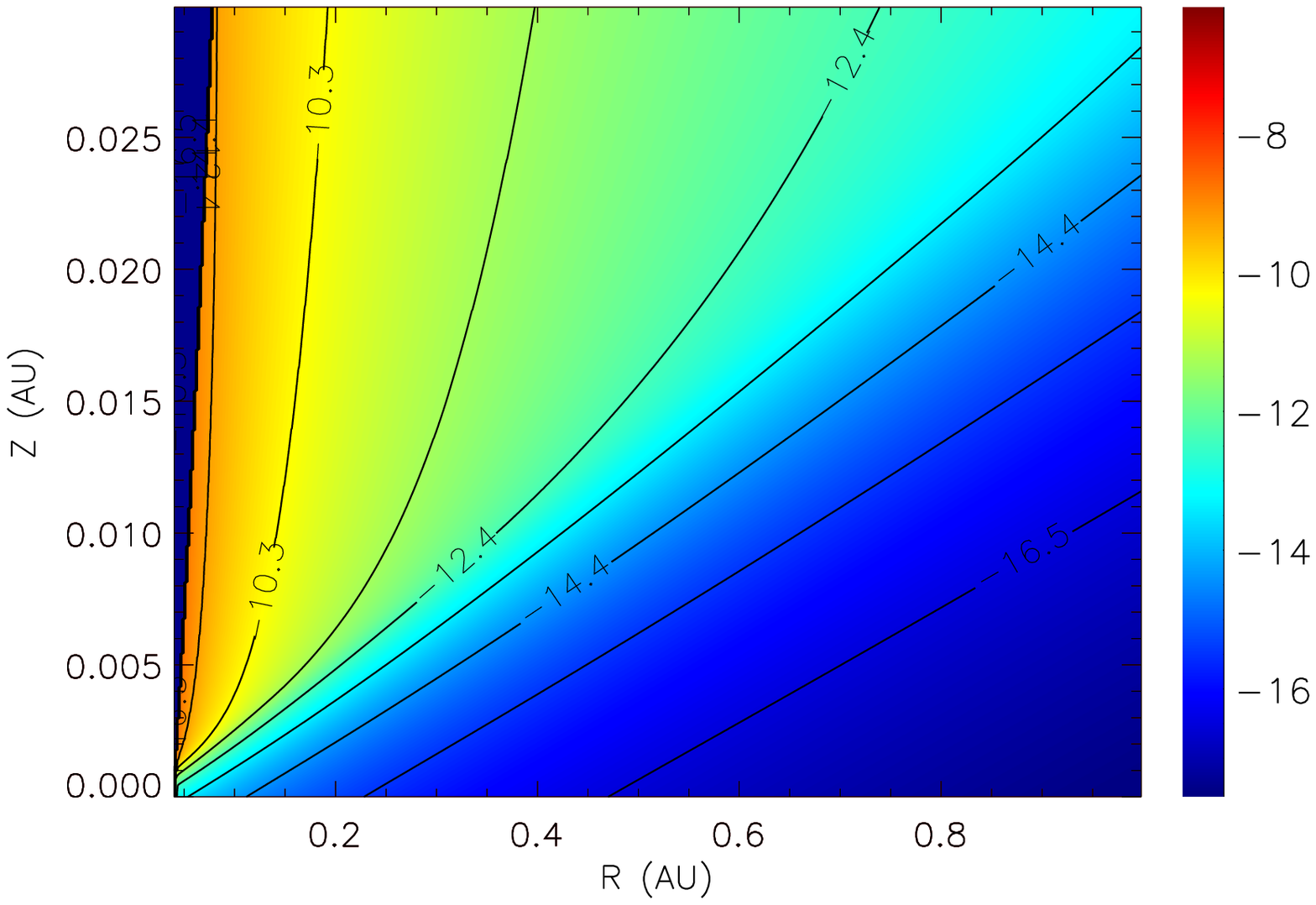}
\caption{Log-ionisation rate (s$^{-1}$) of the self-similar accretion-ejection solution presented in
 fig.~\ref{fig:maes}. The system's characteristics are 
 $M=0.56\;\msol$, $\dot M=10^{-6}\;\msol$~yr$^{-1}$,
 $L_X=10^{29}$~erg~s$^{-1}$ and $kT_x=3$~keV. \emph{Left:} the X-ray source is
 located in $z_{\rm src}=0$. \emph{Right:} the X-ray source is located in
 $z_{\rm src}=R_{\odot}$. Black lines are iso-contours.\label{fig:ioni_rate}} 
\end{center}
\end{figure*}

Maps of the log-ionisation rate of our reference MAES structure are given in
fig.~\ref{fig:ioni_rate} for a X-ray source located in $z_{\rm src}=0$ (left
panel) and $z_{\rm src}=R_\odot$(right panel).
As for the density map, the black area near the vertical axis is the region where
the self-similar solution is not defined.
The ionisation rate  spans over ten orders of magnitude, from
$\sim 10^{-10}$~s$^{-1}$ in the wind to $\sim 10^{-20}$~s$^{-1}$ in the disc.
The importance of the X-ray source location is visible when comparing the
two maps. Moving slightly the source from $z_{\rm src}=0$ to $z_{\rm src}=R_\odot$ is enough to increase the ionisation rate of the disc by $\sim$ two
orders of magnitude. As will be seen later, going to higher altitudes
(reconnection of magnetic loops) has an even more drastic effect.

\subsubsection{Dead zones\label{subsec:deadzone}}

In this section, we compare the disc midplane ionisation rate to the critical
rate given by \eq{eq:zeta_crit}. Note that our aim in doing so is not to
determine with precision the geometry of the dead zone (thickness, extension)
but rather to hint at the possibility for a JED to host one. For that reason,
all the results we present here are located at the disc midplane, which is the
hardest region of the disc to ionise. If
$\zeta_X/\zeta_X^{\rm crit}\gtrsim 1$, the mid-plane is MRI-active and no
dead zone is present anywhere, whereas the opposite case tells us that the disc
may harbour a dead zone, without specifying its thickness. 

\begin{figure*}
\begin{center}
\includegraphics[clip=, width=16cm]{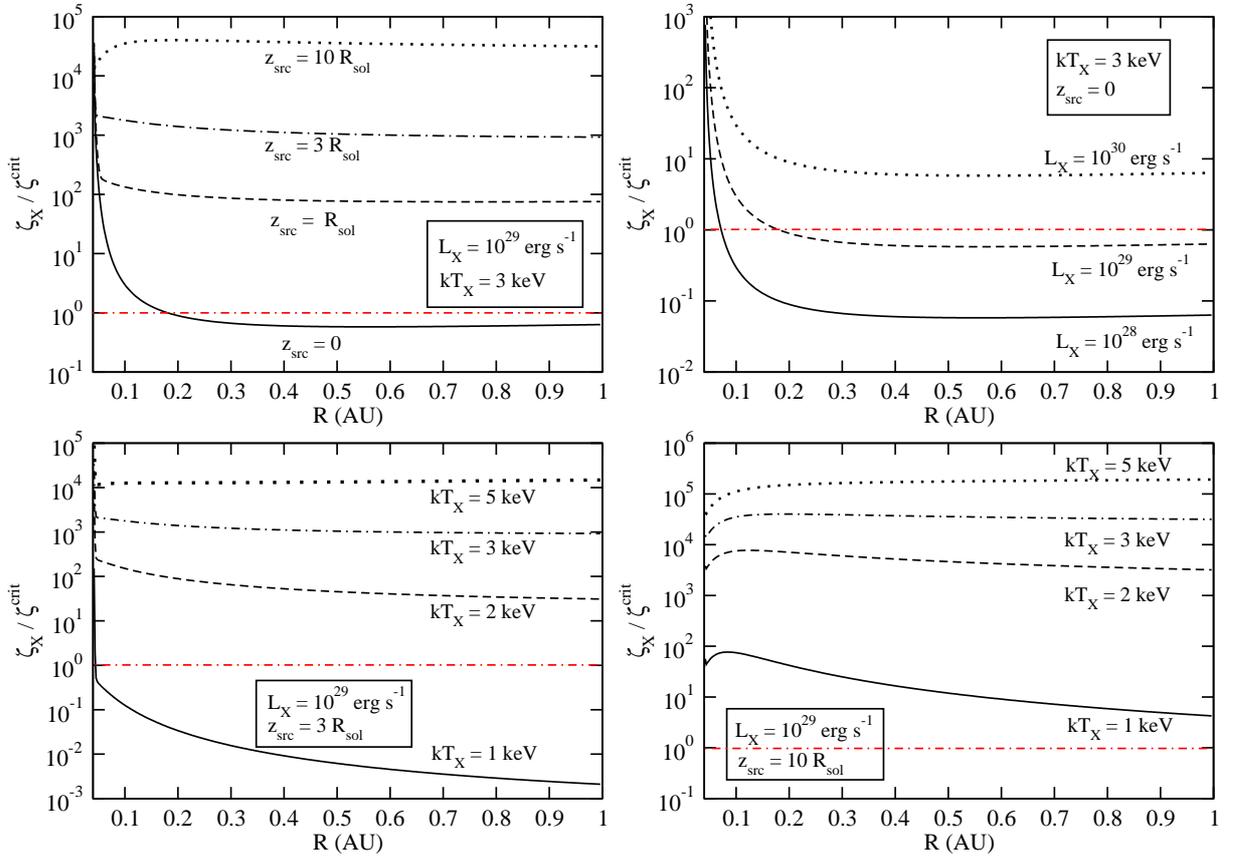}
\caption{Midplane ionisation rate to critical ionisation rate ratio as a
 function of the distance to the protostar. The source location  $z_{\rm src}$ (upper left),
luminosity $L_{\rm X}$(upper right) and photon energy $kT_{\rm X}$ (lower panels) are changed. The
red line corresponds to the critical value of the ratio below which no MRI
can be triggered.\label{fig:compare_ioni_rate}} 
\end{center}
\end{figure*}

The  canonical parameter configuration is $L_X=10^{29}$~erg~s$^{-1}$,
$kT_x=3$~keV and ${\cal R}_m^{\rm crit}=1$. We systematically
explore the parameter space and results are presented in
fig.~\ref{fig:compare_ioni_rate}. Each panel shows the radial variation
of $\zeta_X/\zeta_X^{\rm crit}$ and the red horizontal line represents the critical
value $\zeta_X/\zeta_X^{\rm crit}=1$ (corresponding to ${\cal R}_ m^{\rm crit}=1$). Note
that the critical rate does not depend on the X-ray source
properties and remains constant (at a given radius) throughout: a change in
the $\zeta_X/\zeta_X^{\rm crit}$ ratio translates directly into the same
change in $\zeta_X$.

\paragraph{X-ray source location --- fig.~\ref{fig:compare_ioni_rate} upper
 left panel}
 
As already briefly shown in fig.~\ref{fig:ioni_rate}, the location of the
source w.r.t. the disc midplane is an important parameter. This is made clearer by the upper
left panel of fig.~\ref{fig:compare_ioni_rate} where the ionisation rate
varies over $\gtrsim$ 4 orders of magnitude when the source moves from $z_{\rm
src}=0$ to $z_{\rm src}=10\;R_\odot$ (reconnection in magnetic loop). 
A change in the radial profile is also noticeable due to the change of geometry with the
source acting as a lamp post on the entire disc for the highest altitudes. For all
these configurations, the steady-state hypothesis is valid (see
fig.~\ref{fig:ss_valid}) and the disc appears slightly sub-critical (i.e. with
a possibility of dead zone) only when $z_{\rm src}=0$. But this is inconsistent
with the way we believe these X-rays are actually produced.

\paragraph{Luminosity --- fig.~\ref{fig:compare_ioni_rate} upper
 right panel}

The ionisation rate is directly proportional to the luminosity of the source
and the effect is therefore straightforward. The source is located in $z_{\rm
 src}=0$ and for the lowest luminosity, a significant part of the disc should
harbour a dead zone. However, for this low luminosity, the steady-state
hypothesis is only marginally valid and a time-dependent approach should be 
used to confirmed this result. Nevertheless, even for the lowest luminosity used, the dead zone
disappears as soon as the source is displaced at a slightly higher
altitude (not shown here).

\paragraph{Photon temperature --- fig.~\ref{fig:compare_ioni_rate} lower panels}

The ionisation rate is also very sensitive to the photon temperature. The two
bottom panels in  fig.~\ref{fig:compare_ioni_rate} show the evolution of the
$\zeta_X/\zeta_X^{\rm crit}$ ratio when $kT_X$ is changed between 1 to 5~keV.
First of all, with a source in $z_{\rm src}=3 R_\odot$ (left), the MRI is active in
the entire JED if $kT_X\gtrsim 2$~keV. For lower energies, the disc midplane
appears strongly sub-critical, implying the existence of a dead zone. However,
for this lowest temperature, the \emph{a posteriori} verification showed that
this was a regime where steady-state could not be assumed ($t_{\rm ion}>t_{\rm acc}$) 
and we cannot conclude in the presence of a dead zone without a full time-dependent study.

This timing issue is solved when the X-ray source comes from some
reconnection event ($z_{\rm src}=10\; R_\odot$, right panel). In that case, the ionisation
rate is always larger than its critical value, whatever the energy of the
X-ray photons. No dead zone can be present.

\paragraph{Recombination process}

In order to calculate the ionisation fraction $x_e$, we have assumed that
dissociative recombination with molecular ions was the main mechanism, and 
parametrised it with an effective value $\beta \sim 3\times 10^{-6}T^{-1/2}\;{\rm
 cm}^{3}\;{\rm s}^{-1}$
\citep{1999ApJ...518..848I,2002MNRAS.329...18F,2003ApJ...598..645M}. 
This value actually depends of the composition of the gas, each species 
having its own recombination rate. Theoretical and experimental studies give a 
range of individual parametrisations $\beta=\beta_0 T^{-n}$ 
(with $n$ often found $\sim$ 0.5, \citealp{2006PhR...430..277F}). Using these values, 
several authors have treated the full, or a reduced, chemical reaction network in their dead zone calculation 
(e.g., \citealp{2000ApJ...543..486S,2004A&A...417...93S}), 
but this is beyond the scope of the present paper. 
When metals dominate, radiative rather than collisional recombination occurs, 
in which case $\beta=\beta_r=3\times 10^{-11}T^{-1/2}\;{\rm cm}^{3}\;{\rm s}^{-1}$ 
\citep{2002MNRAS.329...18F,2003ApJ...598..645M}. 

In order to evaluate the sensitivity of our results to the recombination process, 
a simple exercise consists in using a parametric expression for the
recombination coefficient $\beta=\beta_0 T^{-1/2}$ and changing the normalisation factor $\beta_0$ from $3\times 10^{-11}$ to $3\times 10^{-6}\;{\rm
 cm}^{3}\;{\rm s}^{-1}$ . This does not provide us with quantification for a realistic process
but rather gives a feeling on how the situation changes with this
coefficient\footnote{Note that when the recombination coefficient changes, so
 does the critical ionisation rate which is then not given by
 \eq{eq:zeta_crit} anymore.}. For this reason, we do not present a figure but
simply summarise our findings below. For the canonical parameters
$L_X=10^{29}$~erg~s$^{-1}$, $kT_X=3$~keV, it is found that:

\begin{itemize}
\item For a source in $z_{\rm src}=0$, the steady-state hypothesis is valid on
 the entire radius range for $\beta_0 \gtrsim 3 \times 10^{-8}\;{\rm
 cm}^{3}\;{\rm s}^{-1}$, but fails below. 
In this regime, no dead zone is found, whatever  $\beta_0$.
\item If $z_{\rm src}=10\;R_\odot$, there is no restriction to
 the steady-state assumption and the disc midplane is always MRI active.
\end{itemize}

Of course, these conclusions are dependent on the X-ray luminosity, the photon 
energy and source location.  Nevertheless, even for the lowest energies and luminosities, we do not
find any dead zone present in the disc, whatever $\beta_0$, if the X-ray source
comes from a reconnection event in $z_{\rm src}=10\;R_\odot$.

\paragraph{Critical magnetic Reynolds number}

In \S\ref{subsec:critical_ioni_rate}, it was noted that the value of the critical magnetic Reynolds number is still
largely unknown. Figure~\ref{fig:compare_ioni_rate} shows the results obtained for
${\cal R}_ m^{\rm crit}=1$. The critical ionisation rate given by \eq{eq:zeta_crit} scales
with  $({\cal R}_ m^{\rm crit})^2$. All other things being equal, fig.~\ref{fig:compare_ioni_rate}
can also be used to visualise the results obtained for ${\cal R}_ m^{\rm crit}=100$ simply by
shifting the red ``critical'' line from 1 to $100^2=10^4$. This configuration is clearly less favourable to the MRI. 
With $L_X=10^{29}$~erg~s$^{-1}$, only hard flares with $kT_X\gtrsim 3$~keV and located in $z_{\rm src}\gtrsim 7R_\odot$ will provide sufficient ionisation for the MRI to be present in the entire disc. However, these
limits still accommodate most of the X-ray flare properties observed in young stellar objects. We therefore
conclude that, even if ${\cal R}_ m^{\rm crit}=100$ makes it easier for a 
JED to host a dead zone, this nevertheless remains unprobable given the observational 
X-ray properties of YSOs.

\subsection{Dependency on the underlying MAES solution\label{subsec:equiv}}

All previous results on the ionisation structure of JEDs have been obtained with our reference solution. 
Are they strongly dependent on the solution used? 

For the set of "warm" (with additional heat deposition at the disc surface  layers \citealp{2000A&A...353.1115C}) and "cold" (e.g. isothermal magnetic surfaces  \citealp{1997A&A...319..340F}) solutions we have available, we calculate the disc midplane ionisation in the way described above. For a fixed stellar mass, accretion rate and X-ray source we find that:
\begin{itemize}
\item For a given $\epsilon$, all warm solutions give the same disc midplane ionisation degree whatever $\xi$.
\item The same applies to all the cold solutions.
\item For given $\epsilon$ and $\xi$, warm and cold solutions give similar results.
\item For a given $\xi$ but different $\epsilon$, the disc ionisation changes strongly.
\end{itemize}

\begin{figure}
\begin{center}
\includegraphics[clip=,width=\columnwidth]{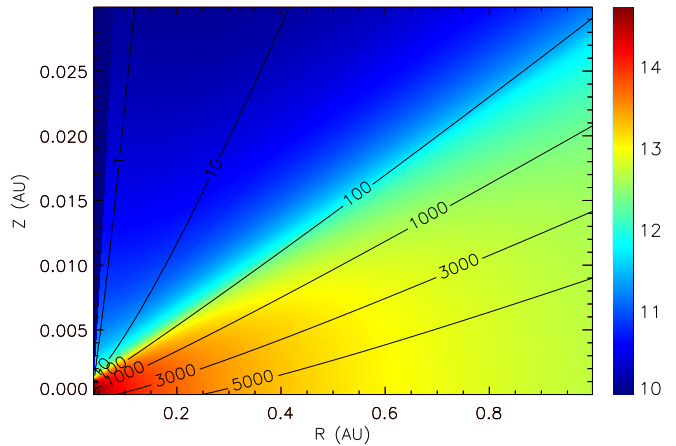}
\caption{Solid black lines: contours of the optical depth $\tau$ for a 1~keV X-ray source located in $z_{\rm src}=0.01$~AU. 
The values of $\tau$, from top to bottom are $\tau=1,\,10,\,100,\,1000,\,3000$ and $5000$. The colour scale represents the log-density (cm$^{-3}$) of our reference
solution, as in fig.~\ref{fig:maes}\label{fig:opt_depth}.}
\end{center}
\end{figure}

These results show that the wind properties, be it through the extra-heating 
term of the warm solutions or the value of the ejection index $\xi$, play a very little 
role on the ionisation state of the disc. We do not show the corresponding plots as 
the almost superimposing lines give no more insight that what has already been said.
The equivalence between cold and warm solutions w.r.t to the ionisation of the disc 
is easily understood as the wind is very tenuous, no matter what the configuration, 
and brings therefore a only a small contribution to the column density between the source and the disc. 
We illustrate the latter point in fig.~\ref{fig:opt_depth}
where the solid contours represent the optical depth $\tau\,(kT_X=1\;\rm{keV})$, for an X-ray source
located in $z_{\rm src}=3R_\odot\sim 0.01$~AU, over-plotted on a density map of our reference solution.
The optical depth achieved in the wind is, at best, a few percent of that in the disc and 
therefore expect the disc ionisation to rely solely on its density distribution, regardless of the properties of its  wind.

For a given accretion rate, the local disc density depends on the amplitude of the accretion speed 
$u_r= m_s c_s = m_s \epsilon V_k$ where $m_s$ is the sonic Mach number (measured here at the disc midplane), $c_s$ the sound speed and 
$V_k$ the Keplerian velocity. {\em The main and major property of a JED is that accretion proceeds roughly at a sonic pace, namely $m_s \sim 1$}. 
This is a consequence of having magnetic fields close to equipartition \citep{1995A&A...295..807F} and explains 
why cold and warm solutions give similar results.

On the other hand,  for a given accretion rate, the disc aspect ratio $\epsilon$ is a critical parameter. The density in the disc scales as $\epsilon^{-2}$ (see Paper~I) and the optical depth increases when $\epsilon$ decreases, resulting in a lower disc midplane ionisation rate. We conclude that the disc thickness in any MAES solution
is the key element to the disc ionisation and that warm or cold solutions with the same aspect ratio are readily interchangeable in that respect.

\subsection{JED ionisation during protostellar evolution}

\begin{figure}
\includegraphics[clip=, width=\columnwidth]{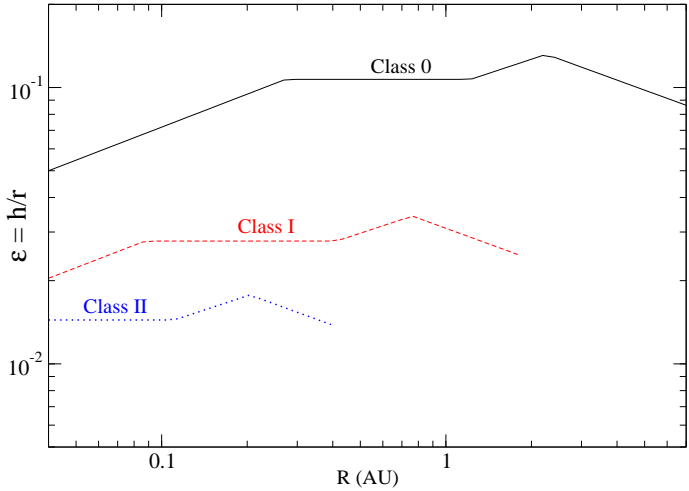}
\caption{The disc aspect ratios obtained for parameters typical of Class~0, I and II
protostars (see text for details), using the approach developed in paper~I. The curves end at the radius beyond
which the JED becomes optically thin.\label{fig:epsilon_classes}}
\end{figure}

So far, the paper has dealt with the prototype of a Class~I 
YSO ($M=0.5$~M$_\odot$, $\dot M_{\rm acc}=10^{-6}$~M$_\odot$~yr$^{-1}$)
for which the JED calculation in paper~I gave a disc aspect ratio 
$\epsilon\sim 0.03$. A \emph{warm} MAES solution having both the 
right value of $\epsilon$ and appropriate Class~I/II jet properties has been 
used to derive the ionisation degree in the JED of such a source.
In order to extend this to the other stages of star formation, namely
Class~0 and Class~II protostars we change $M$ and $\dot M_{\rm acc}$
accordingly and calculate the new values of $\epsilon$ following
the method in paper~I.  As illustrated in fig.~\ref{fig:epsilon_classes}:
\begin{itemize}
\item For a Class~0 object with $M=0.1$~M$_\odot$ and $\dot M_{\rm acc}=10^{-5}$~M$_\odot$~yr$^{-1}$,
we find an average JED aspect ratio of $\epsilon\sim 0.1$.
\item For a Class~II object with $M=0.5$~M$_\odot$ and $\dot M_{\rm acc}=10^{-7}$~M$_\odot$~yr$^{-1}$, we obtain $\epsilon\sim 0.01$.
\end{itemize}
As discussed in \S\ref{subsec:equiv} the disc ionisation state is independent of the jet properties. We therefore use isothermal (\emph{cold}) MAES solutions with these values of $\epsilon$ as they involve no extra physical ingredient (such as additional heat input at the disc surface).

\begin{figure}
\includegraphics[clip=, width=\columnwidth]{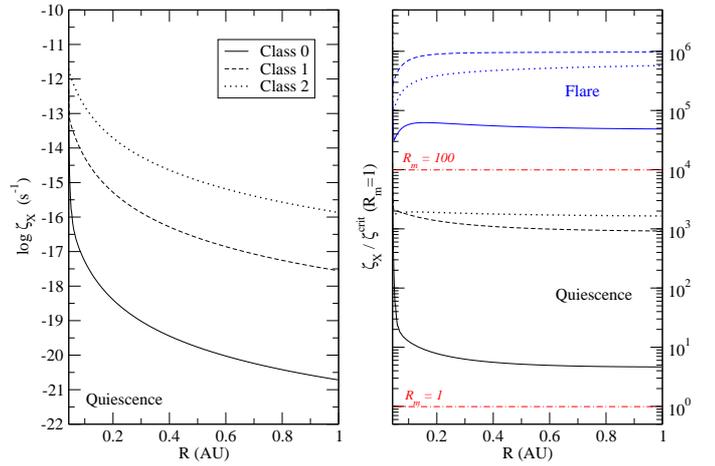}
\caption{\emph{Left:} Log-ionisation rate (s$^{-1}$) on the disc midplane for a typical Class~0 (solid line), Class~I (dashed) and
Class~II (dotted) YSO. X-ray characteristic quiescence properties have been used 
($L_X=10^{29}$~erg~s$^{-1}$, $kT_X=3$~keV, $z_{\rm src}=3~R_\odot$). \emph{Right:} ionisation rate to its critical value for the same 
three star formation stages, calculated for ${\cal R}_m^{\rm crit}=1$. Both quiescence (black) and flaring X-rays (blue, 
$L_X=10^{30}$~erg~s$^{-1}$, $kT_X=4$~keV, $z_{\rm src}=10~R_\odot$) have been considered. The two dash-dotted red lines
correspond to the threshold below which a dead zone is present, whether ${\cal R}_m^{\rm crit}=1$ or ${\cal R}_m^{\rm crit}=100$. 
\label{fig:classes}}
\end{figure}

Figure~\ref{fig:classes} shows the disc midplane ionisation rate (left panel) and 
the ratio of the ionisation rate over its critical value (right panel) for Class~0/I/II YSO prototypes, in solid, dashed and dotted lines respectively. The ionisation rate increases when moving from the youngest to the most
evolved protostars (left). This is easily explained as Class~0 discs are both denser
and thicker than their more evolved counterparts, resulting in a smaller fraction of X-rays
reaching the disc's midplane.

The ratio $\zeta_X/\zeta_X^{\rm crit}$ is shown on the right panel for both quiescent (black)
and flaring (blue) configuration of the X-ray source, calculated for ${\cal R}_m^{\rm crit}=1$. 
The red dash-dotted lines correspond to the dead zone existence threshold for ${\cal R}_m^{\rm crit}=1$
or ${\cal R}_m^{\rm crit}=100$. Typical conservative quiescent 
X-ray properties ($L_X=10^{29}$~erg~s$^{-1}$, $kT_X=3$~keV, $z_{\rm src}=3~R_\odot$) make the entire
disc MRI-active if ${\cal R}_m^{\rm crit}=1$ but inactive if ${\cal R}_m^{\rm crit}=100$, whatever the 
stage of stellar formation; this is especially true for the Class~0 objects. 
However and whatever the threshold, all types of protostars are MRI-active
up to the disc's midplane when considering  X-ray flaring characteristics 
($L_X=10^{30}$~erg~s$^{-1}$, $kT_X=4$~keV, $z_{\rm src}=10~R_\odot$).

In the light of these results it appears that JEDs are very unlikely locations for dead zones to exist,
whatever the evolution of the protostar may be. Class~0 JEDs are more prone to host one than Class~I/II
but this remains marginal and highly dependent on the quiescence X-ray source location and value of 
critical magnetic number.

\section{Discussion and concluding remarks \label{sec:discussion}}

We have revisited the question of the dead zones in accretion discs around young stellar objects by considering a new class of accretion solutions, 
Jet Emitting Discs. A JED is an accretion discs threaded by a large scale magnetic field, giving rise to self-confined jets 
carrying away a significant fraction of the disc angular momentum and energy. The jet emitting disc model relies on the self-similar
MAES solutions, the magnetic fields of which are found smaller than, but close to, equipartition. This is close to the marginal stability 
criterion of the MRI, but still in the regime that allows it to be sustained.

Only a small fraction of the released accretion energy of JED is converted 
into heat. For a given accretion rate, a JED is therefore thinner than a standard accretion disc (SAD). Moreover, because the torque due 
to the jets is dominant, accretion proceeds at a speed close to the speed of sound. This, in turn, leads to discs with a column density much smaller 
than in SADs. This directly translates into a more penetrating X-ray radiation, resulting in a higher ionisation degree compared to that of SADs.
  
The ionisation structure of JEDs has been calculated using self-similar solutions along with the knowledge of their thermal structure. The radial 
extent of a JED is unknown as it depends on the importance of the magnetic flux available in the central regions of the disc. Thus, it certainly 
varies from one object to another. Moreover, for any given object, this extension will probably also vary in time from Class 0 to Class II. In 
this paper, the JED extent has been arbitrarily chosen between $r_{\rm in}=0.04$ AU and $r_{\rm J}=1$ AU only, as it corresponds to the location 
where Class I/II jets seem to be launched: $r_J = 1-3$~AU in the Class~I protostar DG~Tau and $r_J \sim$~ a few 0.1~AU for some Class~II YSO 
\citep{2003ApJ...590L.107A, 2006A&A...453..785F}. With this caveat in mind and given the X-ray properties of young stellar objects, this work 
has mostly ruled out the existence of dead zones in JEDs.

We did not include grain chemistry in our calculation arguing it may not
be relevant to JEDs, as they occupy the innermost regions of the discs. 
Previous studies have also shown gas-only and gas-grain chemical networks to converge in these innermost regions,
comforting our assumption. 
However, the earliest stages of star formation may have JED extending up to a few AU, a distance 
at which grains could have a significant effect. 
Only a model using both gas-phase and gas-grain chemistry, be it a numerical chemical network calculation 
(e.g. \citealp{2004A&A...417...93S}) or a semi-analytical approach \citep{2009ApJ...698.1122O} could settle this issue
and this should be considered in future work.

This result must also be modulated by our ignorance of the critical value of the magnetic Reynolds number above which the disc is MRI-active. Indeed, 
taking ${\cal R}_m^{\rm crit}=1$ allows the entire JED to be active, even during the -- relatively -- soft and low-luminosity quiescent state of YSOs. 
However, if ${\cal R}_m^{\rm crit}=100$, then the critical ionisation degree is only achieved during X-ray flares. In that case, MRI can be continuously 
active in the entire JED only if the recombination timescale is longer than the quiescence phase between two flares, which typically lasts a few days. 
This is easily done for low values of the recombination parameter $\beta_0$ (such as in the metal dominated case where $\beta_0=3\times 10^{-11}$~cm$^3$~s$^{-1}$), 
but becomes more problematic for high values (i.e., $\beta_0=3\times 10^{-6}$~cm$^3$~s$^{-1}$). In this case, we cannot exclude the possibility for a 
dead zone to be intermittently present in
JEDs during the X-ray protostellar quiescent phases and this effect should be stronger in 
the youngest (Class~0) objects. That matter requires a full time-dependent treatment that
is beyond the scope of the present paper.

The absence of a dead zone in the innermost regions of circumstellar accretion discs depends on the outer extent $r_J$ of the JED. The direct
 observation of jet footpoints is not yet within the reach of current instruments. In the near future, ALMA (\emph{Atacama Large Millimeter Array}) 
may be able to directly resolve the launching regions of the closest and most powerful YSO jets, but in the meantime only indirect methods can be 
used. The value of $r_J$ and how it varies in time is an open issue as it depends on the magnetic field transport in discs. Since this transport 
is also dependent on the ionisation properties of the disc, we are facing here a typical "chicken or the egg" dilemma until a self-consistent treatment 
is made. For Class~0/I objects, standard approaches provide an inner dead zone boundary that can be significantly smaller than 1 AU, our typical $r_J$. 
A SAD-JED radial transition could then be operated if the vertical magnetic field reaches a value close to equipartition. Within this framework, an outer dead zone would directly connect to an inner MRI-active JED. This requires of course enough field 
advection in the outermost SAD and through the dead zone. This possibility clearly deserves further investigations (see for instance \citealp{2008ApJ...677.1221R}).

The general trend is nevertheless that Jet Emitting Discs have smaller densities than SADs, sonic accretion and are completely MRI-active: this makes 
them a hostile environment for planet formation. As a consequence, that process must be initiated and largely completed in the outer regions beyond the JED. 
Remarkably, several authors have pointed out the important role that the dead zone, and especially its innermost boundary, plays in both planet 
formation and migration processes. Indeed, solid materials may be trapped at its inner edge leading to the formation of planetary cores \citep{2009ApJ...690..407K}, 
possibly within Rossby vortices 
\citep{2006A&A...446L..13V}. This inner edge is systematically associated with a sudden decrease of the disc column density and disc scale height. 
It thereby provides a very efficient means to stop the inward  type I migration of proto-planetary cores and may be a key ingredient to save planetary 
systems \citep{2006ApJ...642..478M, 2009ApJ...691.1764M}. Note that these effects would be also at work at a SAD-JED transition. 
If planet formation starts early on in the protostellar evolution (Class 0/I), when our results suggest that the dead zone should be truncated 
by a JED, then the conclusions of the above studies, if qualitatively unchanged, should be quantitatively reconsidered.

\begin{acknowledgement}
CC acknowledges a STFC rolling grant. This work was partly supported by the European Community Marie Curie
Actions - Human Resource and Mobility within the JETSET (Jet Simulations,
Experiments and Theory) network under contract MRTN-CT-2004 005592.
The authors also wish to thank Sylvie Cabrit and Paolo Garcia for fruitful discussions.

\end{acknowledgement}


\end{document}